\documentclass[%
 reprint,
superscriptaddress,
 amsmath,amssymb,
 aps,
 pra,
floatfix,
]{revtex4-2}

\usepackage{graphicx}
\usepackage{dcolumn}
\usepackage{amsmath}
\usepackage{bm}

\DeclareMathOperator\sinc{sinc}

\begin{document}

\preprint{APS/123-QED}

\title{Characterizing multiphoton excitation \\using time-resolved X-ray scattering}

\author{Philip H. Bucksbaum}
\affiliation{Department of Physics, Stanford University, Stanford, California 94305, USA}

\affiliation{Department of Applied Physics, Stanford University, Stanford, California 94305, USA}

\affiliation{Stanford PULSE Institute, SLAC National Accelerator Laboratory, Menlo Park, CA 94025, USA}

\author{Matthew R. Ware}
\affiliation{Stanford PULSE Institute, SLAC National Accelerator Laboratory, Menlo Park, CA 94025, USA}

\author{Adi Natan}
\affiliation{Stanford PULSE Institute, SLAC National Accelerator Laboratory, Menlo Park, CA 94025, USA}

\author{James P. Cryan}
\affiliation{Stanford PULSE Institute, SLAC National Accelerator Laboratory, Menlo Park, CA 94025, USA}

\affiliation{Linac Coherent Light Source, SLAC National Accelerator Laboratory, Menlo Park, CA 94025, USA}




\author{James M. Glownia}

\affiliation{Stanford PULSE Institute, SLAC National Accelerator Laboratory, Menlo Park, CA 94025, USA}

\affiliation{Linac Coherent Light Source, SLAC National Accelerator Laboratory, Menlo Park, CA 94025, USA}

\date{\today}
\begin{abstract}
Molecular iodine was photoexcited by a strong 800 nm laser, driving several channels of  multiphoton excitation.
The motion following photoexcitation was probed using time-resolved X-ray scattering, which produces a scattering map
$S(Q,\tau)$.
Temporal Fourier transform methods were employed to obtain a frequency-resolved X-ray scattering  signal
$\tilde{S}(Q,\omega)$.
Taken together,
$S(Q,\tau)$ and
$\tilde{S}(Q,\omega)$ separate different modes of motion, so that mode-specific nuclear oscillatory positions, oscillation amplitudes, directions of motions, and times may be measured accurately. Molecular dissociations likewise have a distinct signature, which may be used to identify both velocities and dissociation time shifts, and also can reveal laser-induced couplings among the molecular potentials.
\end{abstract}

\maketitle


\section{\label{sec:Into}
Introduction}

Intense ultrafast laser irradiation of molecules leads to nonlinear processes that can include multiphoton absorption, impulsive stimulated Raman scattering \cite{Ruhman_Joly_Nelson_1987,cite:RuhmanISRS}, and hyper-Raman fluorescence \cite{Terhune_1965,Ziegler_Hyperraman_1990}.
At still higher intensities molecules undergo bond-softening \cite{Bucksbaum_bondsoftening_1990}, tunnel ionization, charge-resonant enhanced ionization \cite{Bandrauk_CREI_1995}, and multiple ionization \cite{Augst_MultipleIonization_1989}.
All of these phenomena have been studied extensively using the tools of fluorescence, coherent light scattering, and photoelectron spectroscopies; however,
these optical-based methods cannot measure directly the sub-Angstrom displacements and femtosecond motion of the molecular nuclei in response to the light.  Here we employ femtosecond hard X-ray scattering, a new method that fills this gap in our measurement capabilities.

\begin{figure}[htpb]
\includegraphics[width=0.9\linewidth]{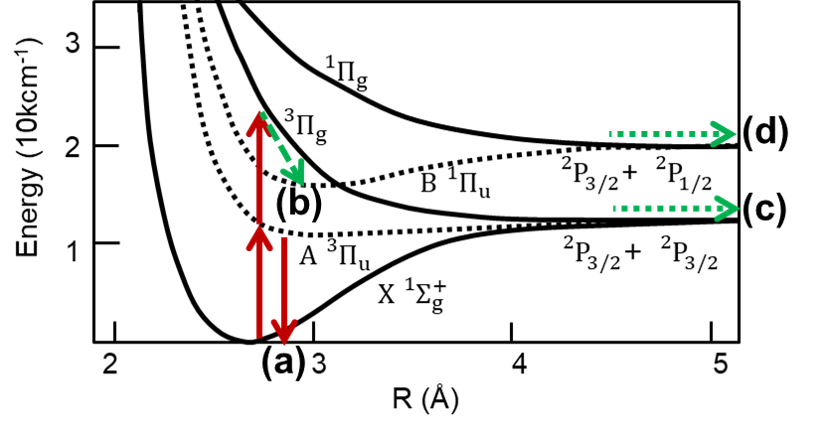}
\caption{\label{fig:I2levels}
The relevant potential energy levels in $I_2$, together with the nonlinear processes observed in this experiment.
(a): Impulsive Simulated Raman Scattering (ISRS).
(b) Spontaneous hyper-Raman scattering (HRS).
(c), (d): Dissociation following 2-photon absorption. }
\end{figure}

We have chosen one of the simplest well-studied systems, molecular iodine irradiated with ultrafast pulses of focused 800 nm radiation
\cite{Schmitt_CARS_I2_1997} (see also Fig. \ref{fig:I2levels}).
This wavelength is resonant with transitions to the weakly-absorbing A state in the molecule.
We work in the impulsive vibrational regime with 70 fs pulses focused to more than $1 \times 10^{13}$ W/cm$^2$.
At this moderately high intensity, both linear and nonlinear photon processes are present in the system.

For example, one prominent nonlinear process excited under these conditions is impulsive stimulated Raman scattering (ISRS)\cite{cite:RuhmanISRS}, a coherent two-photon redistribution of the initial population of the molecule into an oscillating vibrational wave packet in the ground electronic state (Fig. \ref{fig:I2levels}(a)).
This ground-state motion can be tracked through changes to the x-ray scattering pattern
\cite{Haldrup_2019}.
In iodine this Raman cross section is extremely strong.  Furthermore, the ground-state potential energy surface (PES) is nearly parabolic in iodine, so the quantum wave packet generated by ISRS is not highly dispersive and persists for dozens of oscillations.

The experimental setup, data collection, and data preparation used in analysis are all described in section \ref{sec:I2Expt}.
The separate analysis of each nonlinear photo-induced process detected and tracked by the time-resolved X rays is described in section \ref{sec:Analysis}, with sub-sections devoted to each topic.
A table at the conclusion of the paper summarizes each nonlinear phenomenon, its measurement method, and when possible, comparisons to theory.

\section{\label{sec:I2Expt}
Iodine Experiment}

\subsection{\label{sec:TRXS}Time-Resolved x-ray Scattering}

\begin{figure}[htpb]
\includegraphics[width=0.9\linewidth]
{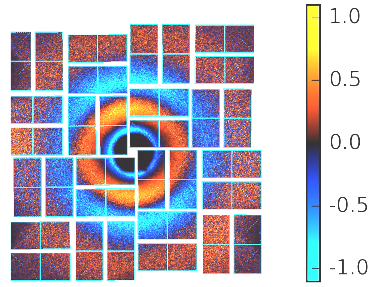}
\caption{\label{fig:40fs} Change in the X-ray scattering distribution recorded for the X-ray pulse delayed by 40 fs following intense laser excitation \cite{Ware_PhilTrans_2019}. Following established procedures, this image has been corrected to remove the polarization-dependence due to Thompson scattering, and to subtract the
background of molecules that were not excited.  Therefore the features seen here are due to the changes in scattering brought about by intense laser excitation.}
\end{figure}

To view the oscillations and other nonlinear laser-induced motion, we employ time-resolved X-ray scattering (TRXS) using femtosecond X rays from the LCLS X-ray free-electron laser \cite{Glownia_prl_2016,Haldrup_2019}.  $1.4$ {\AA} $40~\mathrm{fs}$
X rays scatter from the photoexcited molecules at an adjustable time delay, and the angle of scattering is recorded on an 2-dimensional detector. A typical scattering pattern is shown in Fig. \ref{fig:40fs}.

\begin{figure}[htpb]
\includegraphics[width=0.9\linewidth]{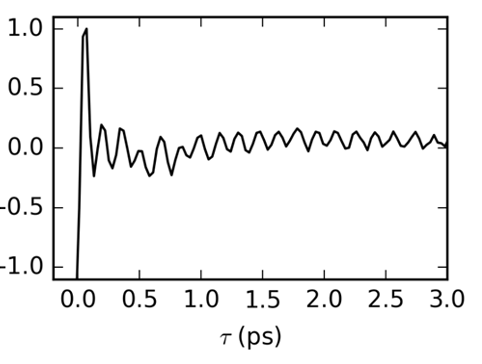}
\caption{\label{fig:OneTile} Signal vs. time delay for a single tile in Fig. \ref{fig:40fs}.  The tile chosen was  the second row, fourth column. }
\end{figure}

\begin{figure}[htpb]
\includegraphics[width=0.9\linewidth]{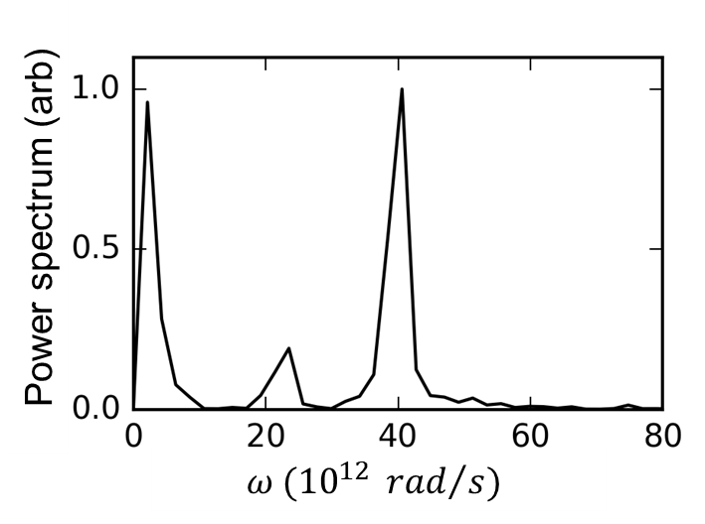}
\caption{\label{fig:1DFFT} Power spectrum (squared amplitude of the Fourier transform) of Fig. \ref{fig:OneTile} }
\end{figure}

Each section of this scattering pattern has its own time dependence.  For example, the integrated signal for one of the tiles shown in Fig. \ref{fig:40fs} oscillates with time delay as shown inf Fig. \ref{fig:OneTile}.  The Fourier transform of this data shows several distinct peaks in Fig. \ref{fig:1DFFT}. Based on our knowledge of the potential energy surfaces in iodine, we can confidently assign the larger peak at $40 \times 10^{12} \mathrm{rad/s}$ to the ISRS process (a) in Fig. \ref{fig:I2levels}, and more tentatively assign the smaller peak at  $25 \times 10^{12} \mathrm{rad/s}$ to process (b) in that figure.  This is the kind and quality of information that has long been available using standard 1-dimensional pump-probe methods such as time-resolved fluorescence. TRXS provides this information for each scattering angle $Q$, and we can use this to obtain information about structure  and motions.

The obvious inversion symmetry in Fig. \ref{fig:40fs}, which is guaranteed by the inversion symmetry of the target gas ensemble, can be used to rewrite the 3-dimensional scattering function $S(\vec{Q},\tau) = S(Q,\cos\theta,\tau)$ as a Legendre series with only even terms $S_{l=0,2,...}(Q,\tau)$,
where
\begin{eqnarray}\label{S_l(Q,tau)}
{
S_{l}(Q,\tau) =
\frac{2l+1}{2}
\nonumber
} \\
{\int_{-1}^{1}
{
P_{l}(\cos \theta)
S(Q,\cos\theta,\tau)
d( \cos\theta)
.}}
\end{eqnarray}

\begin{figure}[htpb]
\includegraphics[width=1.0\linewidth]
{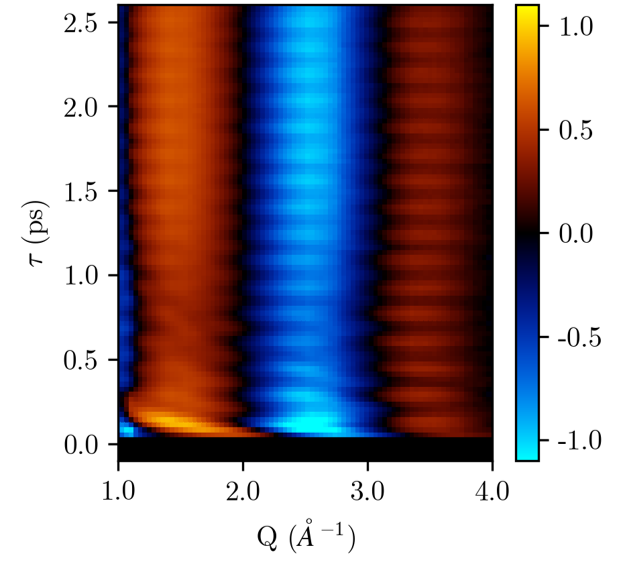}
\caption{\label{fig:S(Qtau)}
$S_0 (Q,\tau)$:
Scattering data projected on to
$P_{0}(\cos\theta)$,
minus unpumped scattering.
}.
\end{figure}

The first two terms in this series, the isotropic term and the term that varies as $P_{2}(\cos\theta)=(3\cos^{2}\theta-1)/2$,
contain most of the data.
Fig. \ref{fig:S(Qtau)}
shows the $l = 0$ isotropic (i.e. angle-integrated) term.  The oscillating pattern that dominates Fig. \ref{fig:OneTile} is still present, but now it can be viewed for every value of $Q$.

\subsection{\label{sec:FRXS}Frequency-Resolved X-ray Scattering}

\begin{figure}[htpb]
\includegraphics[width=1.0\linewidth]
{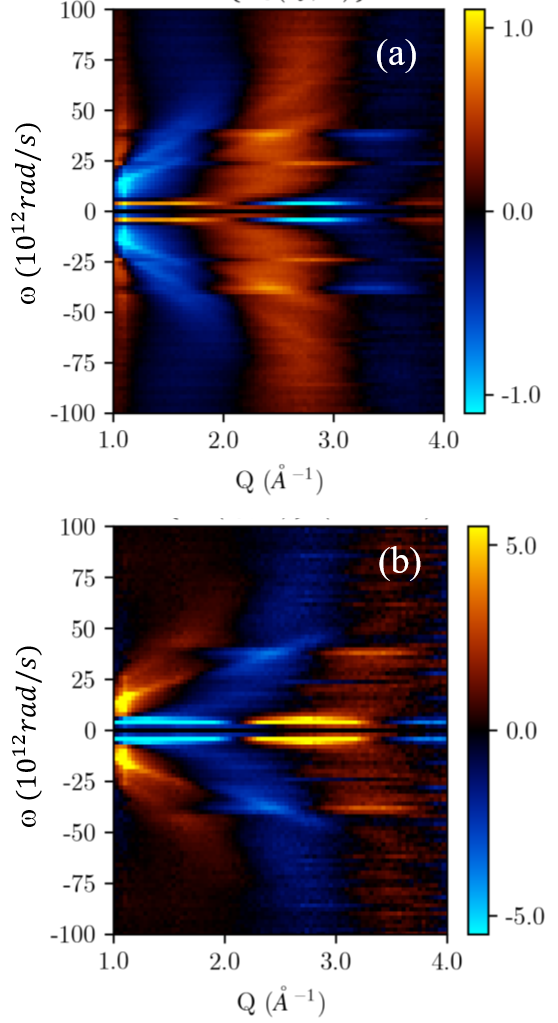}
\caption{\label{fig:S(Qomega)}
Fourier transform of $S_{0}(Q,\tau)$ in Fig. \ref{fig:S(Qtau)} to form
(a) $Re\{\tilde{S}_{0}(Q,\omega)\}$
and
(b) $Im\{\tilde{S}_{0}(Q,\omega)\}$. }
\end{figure}

The same extension to all $Q$ can be performed for the power spectrum in Fig. \ref{fig:1DFFT}.  Here it is useful to retain the full complex Fourier transform $\tilde{S}(Q,\omega)$, shown in Fig. \ref{fig:S(Qomega)}. We will show that the relative phase for different values of $Q$ has physical meaning \cite{Ware_PhilTrans_2019}.

\begin{figure}[htpb]
\includegraphics[width=1.0\linewidth]
{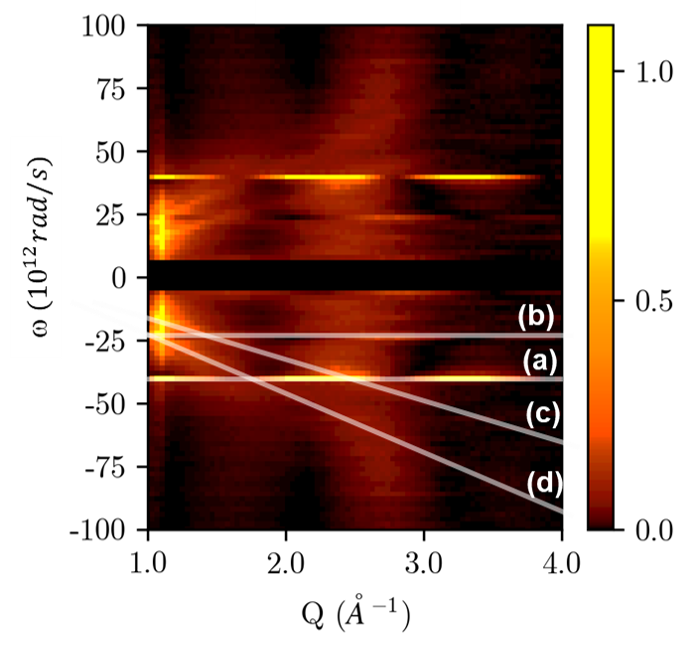}

\caption{\label{fig:Powerspectrum}
The power spectrum $
\mathcal{P}_0 (Q,\omega)=
|
\tilde{S}_{0}(Q,\omega)
|^{2}
$.
The low-frequency part of the plot ($\omega < 5\times 10^{12} \mathrm{rad/s}$) has been darkened to help enhance the contrast.
Nonlinear processes shown in Fig. \ref{fig:I2levels} appear as straight lines when the data is displayed in this way. These are identified in the lower half-plane of the figure:
(a): Impulsive Simulated Raman Scattering.
(b): Spontaneous hyper-Raman scattering.
(c,d): Dissociation.
}
\end{figure}

A real advantage to displaying the data in this way
is that each process shown in Fig. \ref{fig:I2levels} is cleanly separated from the others.  This is even more easily seen in the power spectrum
$
\mathcal{P}_0 (Q,\omega)=
|
\tilde{S}_{0}(Q,\omega)
|^{2}
$
shown in Fig. \ref{fig:Powerspectrum}.
We now consider each of these features in the data to measure the structural information that is associated with each kind of motion.

\section{\label{sec:Analysis}
Analysis}

\subsection{\label{sec:ISRS}Impulsive Stimulated Raman Scattering}

The strongest high-frequency feature in Fig. \ref{fig:Powerspectrum} is the bright narrow horizontal line at $\omega \approx \pm 40 \times 10^{12} \mathrm{rad/s}$.  (The data at angular frequencies below $5 \times 10^{12} \mathrm{rad/s}$ correspond to rotational wave packets that evolve over picoseconds, and will not concern us here.)
This line is associated with ISRS redistribution among the nearly-evenly-spaced vibrational levels of the ground state of the iodine molecule.
A measurement of this feature of Fig. \ref{fig:Powerspectrum} integrated over $Q$
yields a
value of $|\omega|=40.3\pm 1.0 \times 10^{12} \mathrm{rad/s}$, compared to a literature value of $40.42 \times 10^{12} \mathrm{rad/s}$ ($6.43$ THz) \cite{Barrow_Yee_1973}.  Again, this is the same information that can be obtained by all-optical nonlinear spectroscopic methods.

The added advantage of ultrafast X-ray probes of this system is to yield correlated spatial and motion information to accompany the transient spectra.
Data from X-ray scattering can answer several questions about the dynamics of motion in the ISRS process in iodine:
Where is the motion within the molecule that is responsible for the transient spectra?
What is the amplitude of this motion?
What is the phase, i.e. when initially excited, does the molecule first contract or first expand?
Here we will show how these questions may be answered from the data as displayed in Figs. \ref{fig:S(Qtau)}, \ref{fig:S(Qomega)}, and \ref{fig:Powerspectrum}.

\begin{figure}[htpb]
\includegraphics[width=1.0\linewidth]
{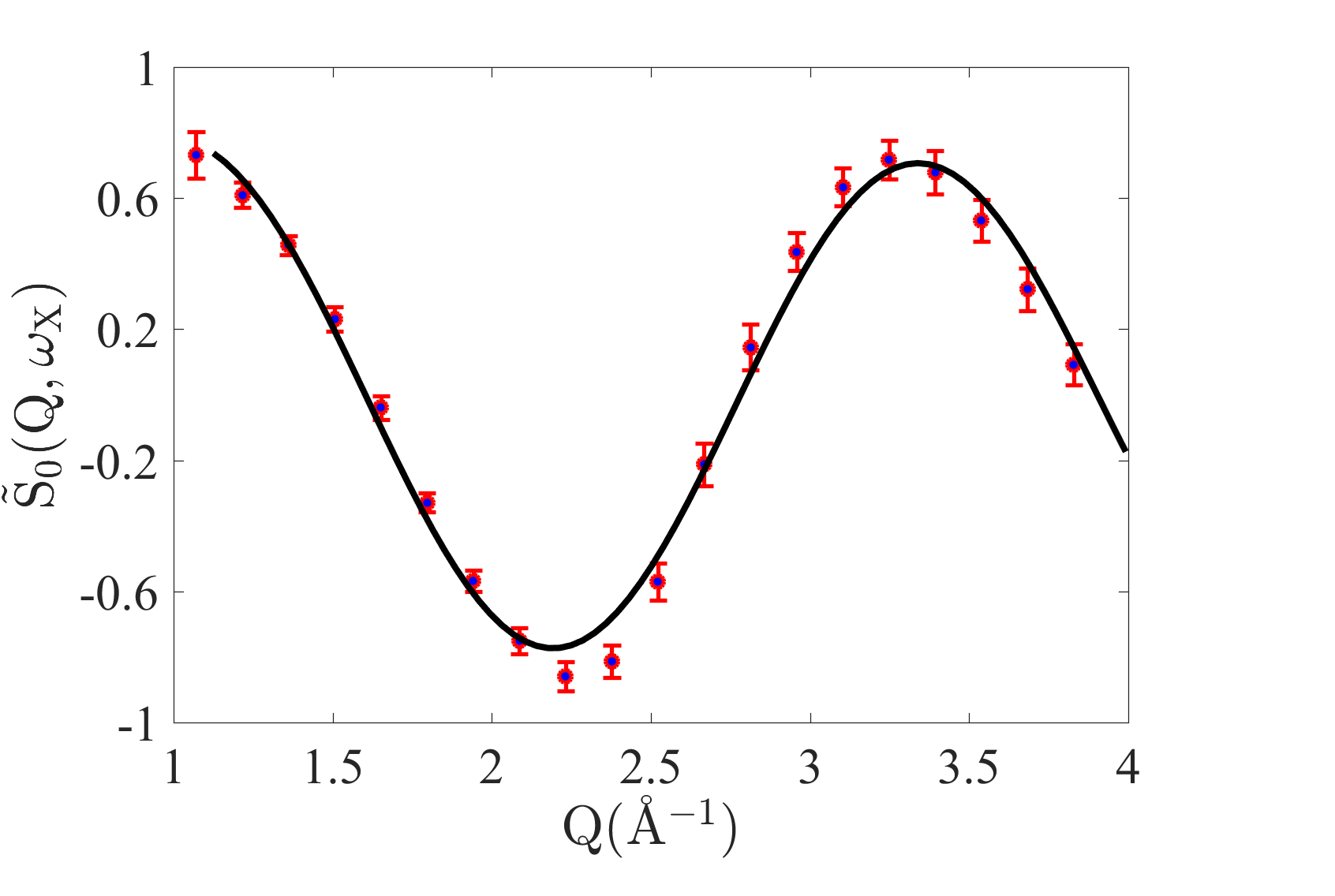}
\caption{\label{fig:40ThzQplot} Points: Data from $\mathrm{Im}\{\tilde{S}_{0}(Q,\omega)\}$ in Fig. \ref{fig:S(Qomega)} in $0.15~\mathrm{\AA^{-1}}$ bins for $\omega = 40.3 \pm 1.0 \times 10^{12} \mathrm{rad/s}$. Line: Fit using Eq. \ref{eq:R0fit}
finds the vibrational mode centered at $R=2.79\pm 0.07 $ {\AA} with vibrational motion amplitude $A= 0.164\pm 0.01$ {\AA}.
For comparison, the iodine ground state has an equilibrium separation of $2.67 $  {\AA}, and the molecular potential well width varies from
$0.173 $  {\AA} at $v=1$ to $0.335 $  {\AA} at $v=5$ \cite{LeRoy_1970}.
}
\end{figure}

The
$Q$-dependence of the ISRS
feature in
Figs. \ref{fig:S(Qomega)} and \ref{fig:Powerspectrum}
reveals the
internuclear separation of the iodine atoms where the
motion occurs.
This can be seen in the lineout displayed in Fig. \ref{fig:40ThzQplot} taken from $\mathrm{Im}\{\tilde{S}_{0}(Q,\omega)\}$.
The scattering signal expected from an ensemble of diatomic molecules with constant internuclear distance $R_0$ is  \cite{Ware_PhilTrans_2019}
\begin{eqnarray}\label{eq:R0fit}
{
\tilde{S}_0(Q,\omega_{0}) = \frac{A}{2R_0}(\cos QR_{0} - \sinc QR_{0})
\delta(\omega - \omega_{0})
}
\end{eqnarray}

This function predicts zeros in the scattered signal at $QR_0 = 0$, $4.5$, $7.7$,  $10.9$, and so on, and these can be used to determine $R_0$ if the amplitude of oscillation $A$ is small.

For finite $A$ but still smaller than $R_{0}$, the scattering pattern of $\tilde{S}(Q,\omega_{0})$ in Fig. \ref{fig:S(Qomega)} and \ref{fig:Powerspectrum} is a similar modulation due to $R_{0}$, but contained within a slowly-varying envelope determined by $A$ \cite{Ware_2019}:
\begin{eqnarray}\label{eq:Afit}
\begin{aligned}
   &\tilde{S}_0(Q,\omega_{0}) \propto
   \\[6pt]
   &\frac{1}{Q}
   \int_0^{Q}{dQ' \sin (Q'R_{0}) J_1 (Q'A)}   .
\end{aligned}
\end{eqnarray}

A fit to
Eq. \ref{eq:Afit} in Fig. \ref{fig:40ThzQplot}
yields a value of
$R_{0} = 2.79 \pm 0.07$  {\AA}, the average interatomic spacing of the ISRS
motion in the laser-excited system.
The expected value for $R_{0}$ depends on the vibrational redistribution brought about by the impulsive Raman interaction, which  strongly favors $\delta v = \pm 1$.
The average iodine atomic spacing for the low-lying vibrational levels is $2.68 $  {\AA}, close to our measured value.

Two features of this measurement are noteworthy: First, we arrived at this value through a \textit{de novo} analysis based on
Eq. \ref{eq:Afit}
without using any prior knowledge or theory of the structure of iodine; and second, this analysis is isolated on the ISRS
motion, and so allows us to filter out any background processes, and even to filter out the portion of our X-ray scattering from iodine molecules that were not excited.

The effect of the finite amplitude of motion is more difficult to discern in this plot. The amplitude creates a spread of $Q$-values, and leads to decay of the ISRS
feature with increasing $Q$.
Our collection geometry restricted the scattering vector $Q$ to less than $4 ~\mathrm{\AA^{-1}}$, so small $A$ leads to a small but  noticeable decay of the oscillation amplitude with $Q$, leading to a best fit value of $A=0.164 \pm 0.01 $  {\AA}.
This fit value is reasonable for a vibrational wave packet in iodine making excursions between the classical turning points in the ground state potential well for $v \approx 4$ \cite{LeRoy_1970}.
This fit, however, assumes that the oscillation amplitude is constant over the period of observation of several picoseconds, and we know that this is not the case.

\begin{figure}[htpb]
\includegraphics[width=1.0\linewidth]
{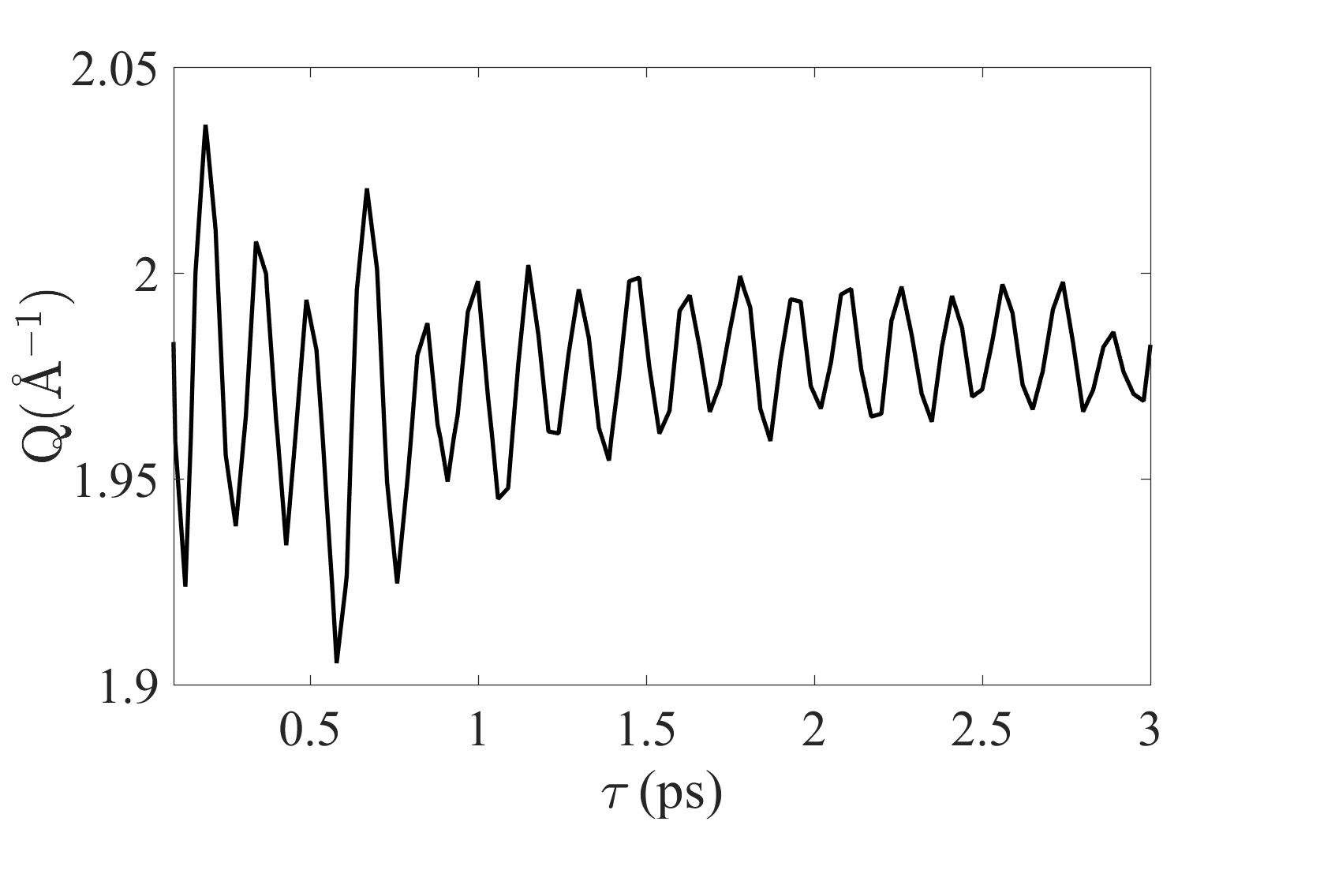}
\caption{
\label{fig:S(Qtau)oscillation}
Oscillation of the node in $S(Q,\tau)$ (Fig. \ref{fig:S(Qtau)}) in the vicinity of 2 {\AA}$^{-1}$.
}
\end{figure}

A more obvious view of this amplitude of oscillation is in the time-domain data in Fig. \ref{fig:S(Qtau)} and \ref{fig:S(Qtau)oscillation}.
As the atomic separation changes, so does the corresponding scattering pattern.
For small amplitude motion, the amplitude of the observed $Q$-space oscillation is approximately linearly proportional to the amplitude of the oscillation in real space, and $\pi$ radians out of phase, since stretching the bond decreases the angle of maximum scattering.
The amplitude of this feature is also proportional to the relative population of oscillating molecules; but if the motion of interest is the dominant mode in the data shown in Fig. \ref{fig:S(Qtau)} (i.e. the excitation fraction to this mode is nearly unity), then the oscillation amplitude can be seen easily in that figure.

Fig. \ref{fig:S(Qtau)oscillation} shows that the scattering pattern $S(Q,\tau)$ in the vicinity of the zero crossing at 2 {\AA}$^{-1}$ oscillates in $Q$ with an rms amplitude that varies from
about 0.05 {\AA}$^{-1}$ near $\tau=0$ to about 0.01 {\AA}$^{-1}$ over 1-3 ps, values that imply oscillation amplitudes smaller than $0.07 $  {\AA}, far below the fit value of Fig. \ref{fig:40ThzQplot}.
This discrepancy is likely due to the fact that the diffraction signal
$S(Q,\tau)$
takes an average over ISRS-excited molecules and a larger background of molecules that were not excited by the laser, while the Fourier transform $\tilde{S}(Q,\omega)$ separates these components cleanly and permits more accurate measurement of the oscillation amplitude.  Furthermore,
The observed rapid decay in the wave packet signal in Fig. \ref{fig:S(Qtau)oscillation} has also been noted in experiments using optical probes and attributed to rotational dispersion of the ensemble molecular alignment \cite{broege_strong-field_2008}. Therefore the value found in the fit to $\tilde{S}(Q,\omega)$ is just a frequency-filtered periodic component of a more complicated non-periodic motion.

The phase of the ISRS oscillation may also be read directly from Fig. \ref{fig:S(Qtau)}, by noting that at $\tau = 0$, the initial oscillation is moving towards \textit{smaller} $Q$.
This shows that immediately following the  ISRS excitation by the laser, the molecule starts to expand.
Another way to recover this information is by comparing the real and imaginary parts of the temporal Fourier transform $\tilde{S}(Q,\omega)$ in Fig. \ref{fig:S(Qomega)}.
This is the Fourier transform of $S(Q,\tau)$, so the phase $\delta$ of an oscillation feature
\begin{eqnarray}\label{eq:oscillation}
Q(\tau)=Q_0 + Q_1 \cos (\omega_0 \tau + \delta)
\end{eqnarray}
is a measure of the starting conditions of motion for that frequency and scattering vector.

For example, a feature oscillating like Eq. \ref{eq:oscillation} with $\delta = 0$ begins its motion at a turning point.  This would appear as a purely real feature centered at $Q_0$ in $\tilde{S}(Q,\omega)$.
The real and imaginary parts of $\tilde{S}(Q,\omega)$ displayed in Fig. \ref{fig:S(Qomega)} show that the $40 \times 10^{12} \mathrm{rad/s}$ feature in
$\tilde{S}(Q,\omega)$ is almost purely imaginary.  We measure $\delta = 290^{o}$ at Q=2.4 {\AA}$^{-1}$. This means that, as we have already seen, $Q$ is near the center of its oscillation and contracting at $\tau = 0$, and therefore the molecule expands at $\tau = 0$.

For completeness we note that the oscillation amplitude and phase can also be measured directly by transforming Fig. \ref{fig:S(Qtau)} to a scattering plot $S(r,\tau)$ of the spatial autocorrelation vs. time, using spatial Fourier transform techniques.  Unfortunately the signal in the appropriate range of $r$ is congested and suffers from poor resolution in these experiments, so this method is difficult in practice.

Information about the motion of a molecule just after
excitation has been inferred from the signal in all-optical
pump-dump protocols, where population is driven
between different potential energy surfaces.
But in such cases the conclusion relies on
knowledge of the potential surfaces in advance.
X-ray scattering does not rely on this prior knowledge,
and can show details of motion purely on the ground state surface.
This may provide more useful connections to collision chemistry.

So far we have only considered the strongest nonlinear excitation in the iodine, ISRS.  Figs. \ref{fig:S(Qomega)} and \ref{fig:Powerspectrum} show several weaker but prominent channels of excitation, which we will now consider.

\subsection{\label{sec:SHRS}Spontaneous Two-Photon Hyper-Raman Scattering (HRS)}

A second and much weaker bound-state oscillation labeled (b) in Fig. \ref{fig:Powerspectrum} is observed at a measured frequency of $\omega_{(b)} = \pm 24 \pm 2 \times 10^{12} \mathrm{rad/s}$ ($3.8 \pm 0.3 ~\mathrm{THz}$). In this section we will show that this feature is due to hyper-Raman scattering to the iodine B state, the process labeled (b) in Fig. \ref{fig:I2levels}.

The sinusoidal variation in $Q$ for this mode has a shorter period than the ISRS mode, and therefore it corresponds to a larger interatomic separation $R_b$.
The amplitude $A_b$ of coherent vibration for this weak mode is not visible in Fig. \ref{fig:S(Qtau)} against the background of the much larger ISRS mode. However, the clear separation between these two processes when the data is displayed as in Figs. \ref{fig:S(Qomega)} and \ref{fig:Powerspectrum} makes it
visible over other backgrounds, particularly for Q > 1.5 {\AA}$^{-1}$. This can be used to extract the vibrational center of motion, its excursion amplitude, and the phase of its oscillation relative to the time of excitation.

\begin{figure}[htpb]
\includegraphics[width=1.0\linewidth]
{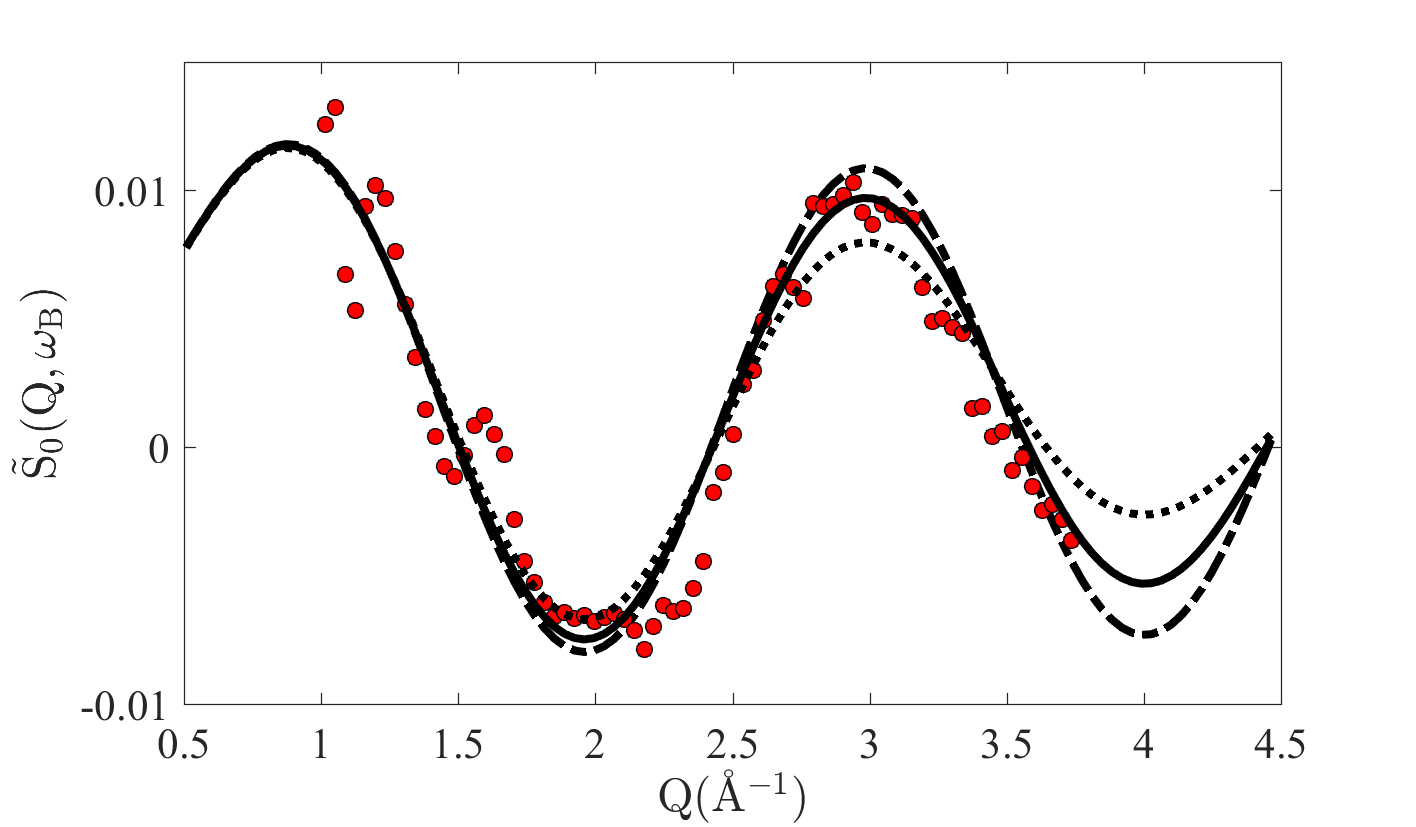}
\caption{\label{fig:S(Qomega_b)}
Dots: $Re(\tilde{S}(Q,\omega_{b}))$ for $\omega_b$ of $24 \pm 2 \times 10^{12} \mathrm{rad/s}$. Solid line: fit to Eq. \ref{eq:Afit} for $A_{b}=0.4 $  {\AA}; Dotted line: fit for $A_{b}=0.6 $  {\AA}.
Dot-dash line: fit for $A_{b}=0.2 $  {\AA}.The fit parameters are the center of oscillation $R_b$, the oscillation excursion $A_b$, the scattering background, which produces the offset from zero, and the overall signal amplitude, which can be compared to the signal from ISRS.
The value of $R_b$ is well constrained by the data: $R_b = 3.10 \pm 0.15 $  {\AA}. The amplitude of oscillation is much less well constrained, since a factor of 2 change in
$A_b$ has similar minimum residuals.  Based on these values we assign this oscillation to coherent vibration in the iodine B state at $v=4$.  This part of the potential well has a width of $0.4 $  {\AA} centered on $R=3.1 $  {\AA}. The oscillation frequency for a superposition of $v=4,3$ is $22.6 \times 10^{12} \mathrm{rad/s}$ ($3.60 ~\mathrm{THz}$) \cite{Barrow_Yee_1973}.}
\end{figure}

According to Eq. \ref{eq:Afit}, valid for $A_b < R_b$, the scattering pattern of $\tilde{S}(Q,\omega_{b})$ in Fig. \ref{fig:S(Qomega)} and \ref{fig:Powerspectrum} is a higher frequency modulation due to $R_{b}$ contained within a slowly-varying envelope determined by $A_b$ \cite{Ware_2019}.
The data $\tilde{S}(Q,\omega_{b})$ and fit to Eq. \ref{eq:Afit} are shown in Fig. \ref{fig:S(Qomega_b)}.
The best fit values are $R_b = 3.10 \pm 0.15 $  {\AA}; and $A_b = 0.4 \pm 0.4 $  {\AA}. The poor precision for $A_b$ is due to the relatively low momentum cutoff for this scattering energy and the lower signal-to-noise ratio of the signal.

Based on these values this feature is consistent with coherent excitation of two or more vibrational states centered at $v=4$ in the iodine B$^{1}\Pi_{u}$ state.  The average spacing of the iodine atoms, the size of the wave-packet motion, and the oscillation frequency all support this conclusion.
The B state lies approximately 2 eV above the iodine ground state, so it is well above resonance for an 800nm photon and well below resonance for two 800 nm photons.  Furthermore, the \textit{ungerade} symmetry of the B state means that this transition can only occur via interactions with an odd number of photons.

We conclude that the three-photon process of hyper-Raman scattering (HRS) (i.e. nonlinear Raman scattering with two photons in and one scattered photon out) is responsible for this excitation
\cite{Cyvin_1965,cite:HRS_Kelley_2010}.  This far off-resonant hyper-Raman scattering occurs during the ultrafast laser pulse, so it can drive vibrational coherences. This process has been studied in other molecular gases and is known to be quite weak, but to our knowledge has never been reported in molecular iodine.  Nonetheless, our ultrafast X-ray data analyzed to construct $\tilde{S}(Q,\omega)$ is capable of filtering out this weak signal, and also providing the amplitude and the phase of the resulting molecular motion.

The comparison of the real and imaginary parts of  $\tilde{S}(Q,\omega_{(b)})$ shows that
the phase of the induced oscillation $\delta$ is consistent with zero, indicating that the coherent oscillation of the molecule immediately after excitation is near its inner turning point, not in the middle of its range like the ISRS
oscillation.  Thus, this weaker oscillation
can be seen in
$\mathrm{Re}\{\tilde{S}(Q,\omega)\}$
but it is barely visible in $\mathrm{Im}\{\tilde{S}(Q,\omega)\}$
in Fig. \ref{fig:S(Qomega)}.

This independent information about the amplitude of oscillation may now be combined with the ratio of signals in Fig. \ref{fig:Powerspectrum} to estimate the branching ratio between ISRS and HRS for this experiment.
The strength of this feature is the product of the fraction of molecules excited to this mode,
times their squared vibrational excursion \cite{Ware_2019}.
Both processes are quadratic in the incident laser fluence, and so many features of the laser focus cancel out of the ratio.  Although the HRS scattered power is weaker by a factor of 8, the squared excursion of motion is larger by a factor of nearly 3, and so the excitation ratio can be estimated to be: $\Gamma_{ISRS}/\Gamma_{HRS} \approx 20:1$.


\subsection{\label{sec:Dissociation}Two-Photon-Induced Dissociation}

We now turn to two other prominent  features in Fig. \ref{fig:Powerspectrum}, which are only barely visible in the standard representation of the data in Fig. \ref{fig:S(Qtau)}: These are the sloped lines labeled (c) and (d), which  increase in frequency magnitude as $Q$ increases.  A straightforward analysis outlined below shows that these features  are due  to scattering from dissociation fragments with with uniform separation velocities (i.e. ballistic trajectories) \cite{Ware_Glownia_Al-Sayyad_O’Neal_Bucksbaum_2019}.

Ballistic dissociation of a pair of molecular fragments that are initially separated by $R_i$ has the form
\begin{eqnarray}\label{eq:R(t)}
{
R(\tau) = R_i + v\tau
}
\end{eqnarray}
where $v$ is the separation speed and $R_i$ is the initial separation assuming ballistic motion.  The projection of the scattering signal associated  motion onto $P_{l = 0}$ is
\begin{eqnarray}\label{eq:S0_dissoc}
{
S_{0}(Q,\tau) =
\int
{
dR R^2 \rho (R,\tau)
j_{0}(QR)
}
}
\end{eqnarray}
where $j_0 (QR) = \sin QR /QR$ is the zeroth-order spherical Bessel function, and
$\rho$ is the nuclear charge distribution.
If we assume for the moment that all the charge is concentrated at a single nuclear separation, i.e. $\rho \propto \delta(R-R(\tau))$, then the temporal Fourier transform of
$S_{0}(Q,\tau)$ contributes a feature to  Fig. \ref{fig:Powerspectrum}
of this form \cite{Ware_Glownia_Al-Sayyad_O’Neal_Bucksbaum_2019}:

\begin{equation}
\label{eq:S_Q_omega_dissoc}
\begin{split}
& \tilde{S}_0 (Q,\omega)
= \int_{0}^{\infty}
{
d\tau \mathrm{e}^{-i\omega \tau}
\frac{sin Q R(\tau)}{QR(\tau)}
}
\\
& =
{
\frac{\mathrm{e}^{i\omega R_i /v}}{2iQv}
}
\times
\\
& {
[E_1 (-i(QR_i -\omega R_i /v))-E_1 (i(QR_i +\omega R_i /v))]
}
\end{split}
\end{equation}

Here $E_1$ is the Exponential Integral function \cite{cite:Abramowitz_and_Stegun}. This function has a sharp maximum when the argument approaches zero, and this means that we expect a peak in Fig. \ref{fig:Powerspectrum} along the line $\omega = Qv$, a line with slope $v$ and intercept $Q=\omega =0$. The two sloped lines seen in each half-plane of Fig. \ref{fig:Powerspectrum} are excellent fits to this form, and we find two values of the speed of separation:

$v_{(c)}=14 \pm 1 $ {\AA}$/ps$

$v_{(d)}=23 \pm 1 $ {\AA}$/ps$

The Iodine molecule is capable of dissociating following excitation by one 800nm photon to the A state, but the separation velocity is slow, about $1.7 $ {\AA}$/ps$, corresponding to a dissociation energy release of $0.04 eV$.  This falls too close the the $\omega$ axis in Fig. \ref{fig:Powerspectrum}, and is also highly dispersed by the shape of the A potential, so it cannot be seen.

Iodine can absorb two 800 nm photons to repulsive \textit{gerade} excited states, which  dissociate to either $^{2}P_{1/2} + {^{2}P_{3/2}}$ or ${^{2}P_{3/2}} + {^{2}P_{3/2}}$ atomic pairs with separation velocities that differ because of the fine-structure splitting in the atomic ground state.
The separation speeds
$v_{3/2+1/2} = 13.4 $ {\AA}$/ps$ corresponds to a kinetic energy release of 1.53 eV. The speed
$v_{3/2+3/2} = 21.6 $ {\AA}$/ps$ corresponds to a kinetic energy release of 0.59 eV.

\subsection{\label{sec:DissocDelay}Measured Dissociation Time Shifts}

These nonlinear dissociation channels could be observed using standard particle detection techniques without the need for X rays; but ultrafast X rays provide additional information that is not readily available in other ways.
This added information comes from the \textit{phase} of the dissociation feature in
$\tilde{S}_0(Q,\omega)$,
clearly visible in Fig. \ref{fig:S(Qomega)}, and also expressed in
Eq. \ref{eq:S_Q_omega_dissoc}.
The quantity in square brackets is purely imaginary when the arguments of the two Exponential Integrals are zero, i.e. at the peak of the dissociation scattering signal.
This cancels the
$i$
in the denominator, so the overall complex phase of this expression is given by the phase of the exponential prefactor, i.e.
$\theta = \omega R_i /v$.
Since this calculation was done for ballistic  motion, we see that
$R_i /v$
is an effective time
$t_i$
which corresponds to the back-extrapolated time when the two atoms with constant speed
$v$
would have separated from a single point at $R_i =0$ if there were no intervening repulsive potential. The value of $t_i$ is referenced to the time when the laser pump and the X-ray probe were coincident.

Neither
$t_i$ nor $R_i$ is zero, of course, for the simple reason that the iodine atoms were not co-located when the laser initiated the dissociation; nor did they travel at a uniform speed at first, since they had to escape the molecular potential.  The time $t_i$ is a kind of ``dissociation time shift,'' a measurable property of the dissociating system, which shows the delay (or advance) suffered by the atoms as they escape the molecular potential. A positive value for $t_i$ indicates that $\tau=0$ occurs after $t_i$, so the molecules are advanced compared  to ballistic motion; a negative $t_i$ means the dissociation is delayed.  Fig. \ref{fig:I2levels} shows that the iodine atoms are separated by $2.7 $  {\AA} when they become excited, and gain most of their kinetic energy in the next Angstrom of travel; so we expect positive values of $t_i$, corresponding to positive dissociation time shifts.

The practical effect of the dissociation time shift can be seen, and also measured, in Fig. \ref{fig:S(Qomega)}.
The scattering associated with dissociation in $\tilde{S}(Q,\omega)$ is real at $Q = \omega = 0$ (out of the range of the measurement), and then the phase advances (or retards) according to the value of $t_i$.
In the data displayed in the figure, the sign of the real part of the scattering signal indicates that the value of $t_i$ is \textit{positive}, which is what we expected.
Furthermore, the second zero appears in the real part of
$\tilde{S}$ at
$Q = 1.8 $  {\AA}$^{-1}$ for dissociation (c), and at
$Q = 2.2 $  {\AA}$^{-1}$ for dissociation (d). This implies that
$\omega R_i / v = Qvt_i = +2 \pi$ at that point.
Together with our measured exit velocities, and combining the errors from these two measurements, we thus recover the dissociation time shifts:

$t_{i(c)} = +250 \pm 20 fs$

$t_{i(d)} = +124 \pm 20 fs$

This is the time advance observed in the dissociating atoms, compared to the time we would measure if they had no interaction and started at the same point.  This time is the physical manifestation of two effects:  First, the atoms start with a physical separation of $R_0 > 0$.  Second, the atoms start nearly at rest, i.e. $v_i = 0 < v$. The net result is the time advance $t_i$.

These dissociation time shifts are  sensitive to the details of the potentials in the molecule. To show this, we compare the measured time shift with times calculated using classical equations of motion and the known potentials in the molecule \cite{Mulliken_1971}.
Dissociation channel (d) has a measured kinetic energy release that shows it must terminate in the production of a pair of ground-state ($^{2}P_{3/2}$) iodine atoms.
This is the more straightforward to analyze, because two-photon absorption from the X state brings the molecule to the vicinity of resonances that correspond to the excitation of the
$(2341) ^{3}\Pi_{g1,2}$ dissociative states, as shown in Fig. \ref{fig:I2levels}.
(Here ``2341'' is shorthand denoting the $\sigma_{g}^{2}\pi_{u}^{3}\pi_{g}^{4}\sigma_{u}^{1}$
molecular orbital configuration.)
There are no other \textit{gerade} states in the vicinity.

Employing Mulliken calculations of these potentials, we find that two-photon resonance occurs at an interatomic separation of
$R=2.73 $  {\AA},
where there is good Franck-Condon overlap with the ground state.  A simple integration of the classical equations of motion on this potential with the atoms starting at rest at
$R=2.73 $  {\AA} yields a dissociation time shift of $t_{i}= +102$ fs, about 20\% smaller than the measurement, but still consistent with it.

Dissociation channel (c), which must terminate in a $^2P_{3/2} ~^2P_{1/2}$ pair based on the kinetic energy released,
is interesting for several reasons. First, there are no \textit{gerade} dissociative states that terminate in this channel with good Franck-Condon overlap at the point of excitation. The  closest is the
$(2341) ^{1}\Pi_{g}$, which comes into resonance at
$R=3.21 $  {\AA}, where the overlap with the ground state is vanishingly small. But even if the molecule did make it to this state to dissociate, the calculated dissociation time shift on this potential and from  this extended separation is $t_{i}= +183$ fs, which is significantly smaller than the measured advance.  If the starting point were $R_0$ (say, for example, if the potential position has been miscalculated by Muliken), this would lead to a still smaller time advance, and so it would not resolve the discrepancy.

This puzzle is compounded by the fact that the upper bound for $t_i$ for any purely repulsive potential  that leads to dissociation to the
$^2P_{3/2}~^2P_{1/2}$ pair
following two-photon absorption from the ground state must be $t_{i} = R_{0}/v_{(c)} = +193$ fs, which is still not advanced as much as we measure.

All of these considerations taken together motivate a different scenario for this dissociation channel.  We propose that the molecules detected in dissociation channel (c) may have been initially excited to the
same Franck-Condon-allowed repulsive potential as those we just described in channel (d).
During dissociation they couple to the \textit{ungerade} B state ($(2431)^{3}\Pi_{u}$), and climb back uphill to the $^2P_{3/2}~^2P_{1/2}$
dissociation channel.  Such \textit{g-u} couplings are nominally forbidden, but can occur weakly due to hyperfine mixing \cite{Pique_Koffend_1984}.  Mixing between \textit{gerade} and \textit{ungerade} states is also facilitated  by the presence of the external laser field \cite{Bucksbaum_bondsoftening_1990}. This pathway has a time advance $t_{i}=+204$ fs, which is larger than the repulsive potential upper bound because part of this path is along an attractive potential.  Though still 18\% smaller than the measurement, this is the closest value of all the possibilities calculated here.

\begin{table*}
\caption{\label{tab:Summary}This a summary of all of the multiphoton absorption and scattering measurements described in this paper.}
\begin{ruledtabular}
\begin{tabular}{ccccc}
Process\footnote{(a), (b), etc. refer to Fig. \ref{fig:I2levels} and Fig. \ref{fig:Powerspectrum}.}
&Measurement&Value measured
&Calculation or expectation\\
\hline
(a) ISRS on the X state
&Center of oscillation
&$2.79\pm 0.07 $  {\AA}
&$2.67$  {\AA} for $v=1\rightarrow v=2$ \cite{Howard_Andrews_raman_1974}\\
&Oscillation frequency $\omega$
&$40.3\pm 1.0 \times 10^{12} \mathrm{rad/s}$
&$40.42 \times 10^{12} \mathrm{rad/s}$ \cite{Barrow_Yee_1973}\\
&Oscillation amplitude
&$0.164\pm 0.01$  {\AA} 
&$0.173$  {\AA} for $v=1$ \cite{LeRoy_1970}\\
&Oscillation phase $\delta$
&$290^{o}$
&$270^{o}$ (Center of oscillation, stretching)\\
\hline
(b) $X\rightarrow B$
&Center of oscillation
&$3.10 \pm 0.15 $  {\AA}
&$3.06 $ {\AA} v=4
\cite{Barrow_Yee_1973}\\
Hyper-Raman scattering (HRS)
&Oscillation frequency $\omega$
&$24\pm 2 \times 10^{12} \mathrm{rad/s}$
&$22.6 \times 10^{12} \mathrm{rad/s}$ for $v=4,3$
\cite{Barrow_Yee_1973}\\
(2-photon inelastic scattering)
&Oscillation amplitude
&$0.4 \pm 0.4 $  {\AA}
&$0.4 $  {\AA} for $v=4$
\cite{Barrow_Yee_1973}\\
&Oscillation phase $\delta$
&$0^{o}$
&$0^{o}$ (purely displacive)\\
&$\Gamma_{HRS}/\Gamma_{ISRS}$
&$5\%$
&\\
\hline
(d) Two-photon dissociation
&Separation velocity $v_{(d)}$
&$23\pm 1$ {\AA}$/ps$
&$21.6$ {\AA}$/ps$\\
to the $^2 P_{3/2} - ^2 P_{3/2}$ state
&Dissociation Time Shift $\tau_{i(d)}$
&$+124\pm 20 \mathrm{fs}$
&$+102 \mathrm{fs}$ for $^3\Pi_g$ dissociation
from \cite{Mulliken_1971}
\\
\hline
(c) Laser-assisted dissociation
&Separation velocity $v_{(c)}$
&$14\pm 1$ {\AA}$/ps$
&$13.4$ {\AA}$/ps$\\
to the $^2 P_{3/2} - ^2 P_{1/2}$ state
&Dissociation time shift $\tau_{i(c)}$
&$+250\pm 20 \mathrm{fs}$
&$+204 \mathrm{fs}$,$^3\Pi_g - ^3\Pi_u$ dissociation
\cite{Mulliken_1971}\\
\end{tabular}
\end{ruledtabular}
\end{table*}

\section{\label{sec:Conclusion}
Summary and Conclusion}

This paper explored ultrafast X-ray detection of two-photon processes in molecules, including impulsive stimulated Raman scattering, hyper-Raman scattering, two-photon photodissociation, and laser-assisted dissociation.  Results are summarized in Table \ref{tab:Summary}.

Nonlinear photoabsorption in molecules produce many effects that can be accurately measured using the new approach of ultrafast hard X-ray scattering, which has been made possible by the current generation of X-ray free-electron lasers.  Some of these phenomena are not detectable at all using all-optical probes.  Ultrafast X rays probes provide \textit{de novo} information about relative atomic positions, velocities, and accelerations, and so these measurements can reveal information about interatomic forces that can lead to insights to aid both theory and experiments.

This research is
supported through the Stanford PULSE Institute,
SLAC National Accelerator Laboratory by the
U.S. Department of Energy, Office of Basic Energy Sciences,
Division of Chemical, Geological and Bio\-sciences,
Atomic, Molecular, and Optical Science Program.
Use of the Linac Coherent Light Source (LCLS),
SLAC National Accelerator Laboratory, is
supported by the U.S. Department of Energy,
Office of Basic Energy Sciences under
Contract No. DE-AC02-76SF00515.
PHB and MRW contributed equally to this work.

\bibliography{apssamp}

\end{document}